\documentclass[fleqn,usenatbib]{mnras}

\usepackage{newtxtext,newtxmath}

\usepackage[T1]{fontenc}
\usepackage{ae,aecompl}

\usepackage{graphicx}	
\usepackage{amsmath}	

\usepackage{color}

\title[SVE bar]{Local variations of the Stellar Velocity Ellipsoid-II: the effect of the bar in the inner regions of Auriga galaxies}

\author[D. Walo-Mart\'in et al.]{Daniel Walo-Mart\'in$^{1,2}$\thanks{E-mail: dwalo@iac.es},
Francesca Pinna$^{3}$,
Robert J.J. Grand$^{1,2}$,
Isabel P\'erez$^{4,5}$,\and
Jes\'us Falc\'on-Barroso$^{1,2}$, 
Francesca Fragkoudi$^{6}$,
Marie Martig$^{7}$
\\
$^{1}$Instituto de Astrof\'isica de Canarias, Calle V\'ia L\'actea s/n, E-38205 La Laguna, Tenerife, Spain\\
$^{2}$Departamento de Astrof\'isica, Universidad de La Laguna, Av. del Astrof\'isico Francisco S\'anchez s/n, E-38206, La Laguna, Tenerife, Spain\\
$^{3}$Max-Planck-Institut für Astronomie, Konigstuhl 17, 69117, Heidelberg, Germany\\
$^{4}$Departamento de F\'isica Te\'orica y del Cosmos, Universidad de Granada, Facultad de Ciencias (Edificio Mecenas), E-18071, Granada, Spain\\
$^{5}$Instituto Carlos I de F\'isica Te\'orica y Computaci\'on\\
$^{6}$Max-Planck-Institut für Astrophysik, Karl-Schwarzschild-Str 1, D-85748 Garching, Germany\\
$^{7}$Astrophysics Research Institute, Liverpool John Moores University, 146 Brownlow Hill, Liverpool L3 5RF, UK\\
}

\date{Accepted XXX. Received YYY; in original form ZZZ}

\pubyear{2020}

\begin{document}
\label{firstpage}
\pagerange{\pageref{firstpage}--\pageref{lastpage}}
\maketitle

\begin{abstract}

Theoretical works have shown that off-plane motions of bars can heat stars in the vertical direction during buckling but is not clear how do they affect the rest of components of the Stellar Velocity Ellipsoid (SVE). We study the 2D spatial distribution of the vertical, $\sigma_{z}$, azimuthal, $\sigma_{\phi}$ and radial, $\sigma_{r}$ velocity dispersions in the inner regions of Auriga galaxies, a set of high-resolution magneto-hydrodynamical cosmological zoom-in simulations, to unveil the influence of the bar on the stellar kinematics. $\sigma_{z}$ and $\sigma_{\phi}$ maps exhibit non-axisymmetric features that closely match the bar light distribution with low $\sigma$ regions along the bar major axis and high values in the perpendicular direction. On the other hand, $\sigma_{r}$ velocity dispersion maps present more axisymmetric distributions. We show that isophotal profile differences best capture the impact of the bar on the three SVE components providing strong correlations with bar morphology proxies although there is no relation with individual $\sigma$. Time evolution analysis shows that these differences are a consequence of the bar formation and that they tightly coevolve with the strength of the bar. We discuss the presence of different behaviours of $\sigma_{z}$ and its connection with observations. This work helps us understand the intrinsic $\sigma$ distribution and motivates the use of isophotal profiles as a mean to quantify the effect of bars.


\end{abstract}





\section{Introduction}
\label{sec:Introduction}
Since the first  extensive imaging surveys it has been shown that bars are common structure in the Local Universe that populate almost two thirds of disc galaxies \citep[e.g][]{2000ApJ...529...93K,2000AJ....119..536E,2007ApJ...657..790M,2009A&A...495..491A,2009MNRAS.393.1531G,2011MNRAS.411.2026M,2015ApJS..217...32B,2018MNRAS.474.5372E}. This fraction decreases with redshifts and at z= 1 only 20\% of observed galaxies host a bar \citep{2008ApJ...675.1141S,2014MNRAS.438.2882M}, indicating that bars play a significant role in the secular evolution of galaxies at later times \citep[see ][for a review]{2004ARA&A..42..603K,2013seg..book..305A}. 

Theoretical studies in the last thirty years have provided valuable information about the formation, stellar orbit distribution and influence of bars in multiple galaxy properties  \citep[e.g.][]{1981A&A....96..164C,1992MNRAS.259..328A,1992MNRAS.259..345A,2000ApJ...543..704D,2003MNRAS.341.1179A,2005ApJ...626..159B,2006ApJ...637..214M,2010ApJ...722..112M,2012MNRAS.427.1429W,2016MNRAS.462L..41F}. Torques induced by bars can efficiently redistribute mass and angular momentum, driving the formation of outer rings \citep{1986ApJS...61..609B,1996FCPh...17...95B,2000A&A...362..465R}, nuclear discs\citep{2006ApJ...645..209D,2014RvMP...86....1S} and inner bars \citep{2012MNRAS.420.1092D,2019MNRAS.484..665D}. Further, bars can also induce the formation of vertically extended bulges, so called boxy/peanut bulges, trough buckling instabilities \citep{1995ApJ...447L..87M,2006ApJ...637..214M,2017A&A...606A..47F}.

Observational studies have investigated the morphology, kinematics and stellar populations properties of bars in profound detail to expand our knowledge on the role that bars play on the secular evolution of galaxies \citep{1986MNRAS.221P..41H,1997A&A...326..449M,2004ARA&A..42..603K,2007A&A...465L...9P,2009A&A...495..775P,2011A&A...529A..64P,2011MNRAS.415.3308G,2011MNRAS.415..709S,2016MNRAS.460.3784S,2019MNRAS.488L...6F,2020A&A...637A..56N}.In particular, \cite{2015MNRAS.451..936S} confirmed that stronger bars have enhanced influence on inner kinematic features (e.g. stronger humps in the velocity profiles and $\sigma$-drops) as predicted by numerical simulations \citep{2003MNRAS.341.1179A,2005MNRAS.364L..18B,2012ApJ...747...60K,2012ApJ...751..124K,2014MNRAS.445.1339L}.  

The information derived from external galaxies rely on projected quantities such as the Line-of-Sight (LoS) velocity distribution from which we can infer  different parameters like the mean velocity, V, and the velocity dispersion, $\sigma$, using both parametric \citep{1993ApJ...407..525V,1993MNRAS.264..712K,2004PASP..116..138C} and non parametric techniques \citep{2000AJ....119.1157G,2014MNRAS.441.2212F,2021A&A...646A..31F}. However, it is difficult to disentangle the multiple families of orbits that coexist in a galaxy from the projected velocities alone, particularly in the inner regions where different structures may coexist with the bar (e.g. b/p bulge, inner bar, nuclear disc). Therefore, understanding the dynamical state of galaxies in terms of their intrinsic components is a key factor to comprehend observations.
In other words, we require a 3D characterization of the dynamical state of galaxies which is typically described through the Stellar Velocity Ellipsoid (SVE)  and its three velocity dispersion components, the vertical, $\sigma_{z}$, radial, $\sigma_{r}$ and azimuthal $\sigma_{\phi}$ velocity dispersions.

Unfortunately, these kind of detailed analysis can only be performed in our Galaxy, where thanks to missions like GAIA \citep{2016A&A...595A...1G} and APOGEE \citep{2017AJ....154...94M}  we have access to the full 6D information \citep[positions and velocities, see the works of][]{2015MNRAS.452..956B,2019MNRAS.489..176M,2021MNRAS.502.1740S}. 
A number of methods have been proposed to study the SVE of external galaxies where the stellar populations cannot be resolved either using analytic models for the disc and velocity dispersion with photometric observations \citep{1999A&A...352..129V}, spectroscopic decomposition of the line-of-sight velocity distribution \citep{2007MNRAS.379..418C,1999MNRAS.303..495E,2006MNRAS.371.1269T,2016MNRAS.460.2720K,2012MNRAS.423.2726G}, or dynamical models \citep{2007MNRAS.379..418C,1999MNRAS.303..495E,2006MNRAS.371.1269T,2016MNRAS.460.2720K}.
Luckily, state-of-the-art hydrodynamical simulations are now able to reproduce realistic disc galaxies with smaller bulges and extended disc thanks to advances in resolution and sub-grid physics modules \citep{2011ApJ...728...51B,2013MNRAS.428..129S,2014MNRAS.437.1750M,2016ASSL..418..317B}, which has lead to an increasing number of studies on the role of bars in galaxy evolution. Simulations of galaxies in the Milky-Way mass range have particularly expanded in the last decade to understand properties of both our Galaxy and external systems, e.g. the Auriga \citep{AurigaProject}, NIHAO-UHD \citep{2020MNRAS.491.3461B} and FIRE \citep{2014MNRAS.445..581H} simulations.

 In this work we aim to study the local properties of the SVE in the inner regions of Milky-Way mass galaxies. To that end we make use of the Auriga galaxies, a set of 30 zoom-in magneto-hydrodynamical cosmological simulations of late-type galaxies. These simulations present different morphologies and evolutionary histories, and  successfully reproduce many present day observables such as stellar masses, sizes, rotation curves, star formation rates \citep{AurigaProject}. Thus, the Auriga sample is the ideal testbed because it will allow us to analyze the impact of external agents and different secular evolution processes in the SVE of high resolution disc galaxies in a cosmological environment. 
 
This paper is the second in a series exploring the spatial variations of the SVE in simulated late-type galaxies. Here we focus on the connection between the SVE and the presence of bars in the inner regions of galaxies. This paper is organized as follows. In Sec.\ref{sec:Methods} we describe the Auriga simulations, our sample selection and the methodology used to study the SVE. In Sec.\ref{sec:first_results} we present the SVE 2D maps and profiles of Auriga galaxies. Sec. \ref{sec:correlation} explores the correlations of the velocity dispersion differences between the major and minor bar axis and different bar morphology proxies. In Sec.\ref{sec:time_evolution} we analyze the time evolution of the velocity dispersion differences and the bar formation. We discuss our results in Sec. \ref{sec:discussion}. Finally, in Sec. \ref{sec:conclusions} we sum up the main results and conclusions from this work.

\section{Methodology}
\label{sec:Methods}

\begin{figure*}
\centering
\includegraphics[width=0.99\textwidth]{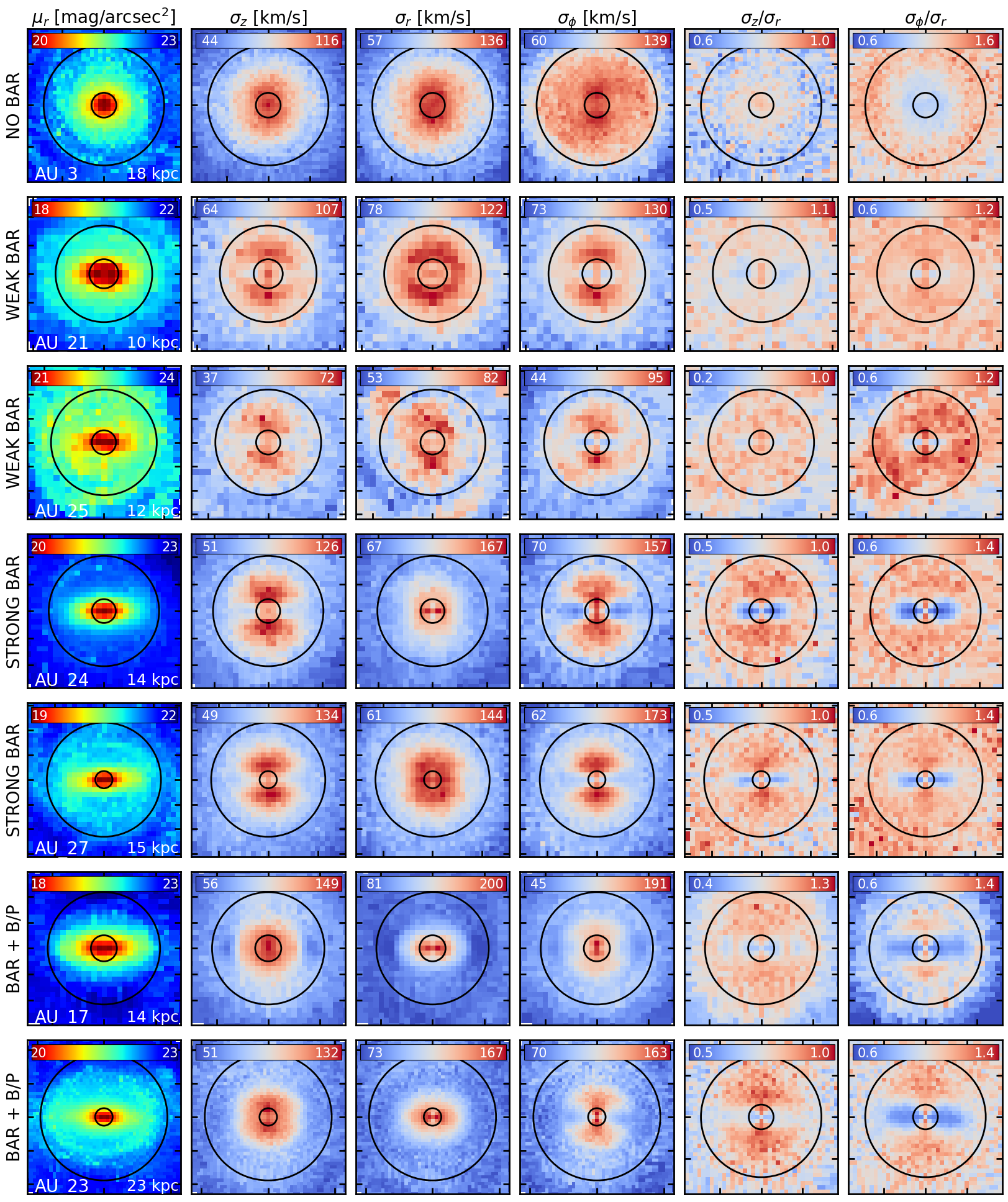}
\caption{The panels show from left to right the surface brightness, SVE velocity dispersions ($\sigma_{z}$, $\sigma_{r}$ and $\sigma_{\phi}$ ) and axial ratio maps ($\sigma_{z}/\sigma_{r}$ and $\sigma_{\phi}/\sigma_{r}$) of 7 galaxies in the Auriga simulations. From top to bottom each row shows the maps of an unbarred galaxy (AU- 3), two galaxies with weak bar (AU-21 and AU-25), two galaxies with strong bars (AU-24 and AU-27) and two galaxies with B/P (AU-17 and AU-23). Inner and outer circles indicate the effective radius of the bulge and the beginning of the disc dominated region. The area covered in this analysis is indicated (in kpc) at the bottom right of the left-most maps of each case.}
\label{fig:FIG1}
\end{figure*}

\subsection{The Auriga simulations}
\label{sec:Auriga}

The Auriga project \citep{AurigaProject} consists of 30 zoom-in high resolution simulations of Milky-Way mass haloes taken from the largest volume Dark Matter Only simulation of EAGLE, L100N1504 \citep[see][for details]{2015MNRAS.446..521S}. These simulations were performed with AREPO \citep{2010ARA&A..48..391S}, an N-body, magnetohydrodynamics (MHD) code. We name each simulation by 'AU-N' where N ranges between 1 and 30. Simulations adopt the cosmological values taken from \cite{2014A&A...571A..16P} ($\Omega_{m}=0.307$, $\Omega_{b}=0.048$, $\Omega_{\Lambda}=0.693$ and $H_{0} = 100 \ \mathrm{h} \ km s^{-1} \ Mpc^{-1}$, with $\mathrm{h} = 0.6777$). In this work, we study simulations where star and dark matter particles have a typical mass of $m_{b}\sim 5 \times 10^{4} M_{\odot}$ of  and $m_{DM}\sim 3 \times 10^{5} M_{\odot}$, respectively. These values correspond to the standard level-4 mass resolution of the Auriga galaxies \citep[see Table 2 in][to find the differences between each resolution level]{AurigaProject}. The comoving softening length for star and dark matter particles is set to $ 500 \mathrm{h}^{-1} cpc$ before z=1, when the simulation adopts a physical gravitational softening length of 369 pc. Groups and subhaloes  were identified by the friend-of-friends \citep{1985ApJ...292..371D} and SUBFIND \citep{2001MNRAS.328..726S} algorithms, respectively, while merger trees were constructed with the LHaloTree algorithm described in \cite{2005Natur.435..629S}. To account for physical processes that act below the resolution limit of the simulation the Auriga models include an updated version of the sub-grid modules used in \cite{2014MNRAS.437.1750M,2014MNRAS.442.3745M}.  For a complete description of the models and the modifications introduced we refer the reader to \cite{AurigaProject}. 

Despite being all of the galaxies in the Milky-Way mass range and their limited number, the Auriga sample presents different evolutionary histories and exhibit a wide range of morphological properties. In terms of morphology, the Hubble-Type of the galaxies ranges  from Sa to Sb and individual structures such as bars, disc breaks, pseudobulges and B/P bulges have been analyzed in detail \citep{2019MNRAS.489.5742G, 2020MNRAS.491.1800B,2020MNRAS.494.5936F, 2021A&A...650L..16F}. In particular, \cite{2020MNRAS.491.1800B} extensively studied the structural properties of the bar such as bar strength, $\rm{f_{bar}}$, length, $\rm{L_{bar}}$ or bar to total (B/T) light fraction, based on mock r-band images. \cite{2020MNRAS.494.5936F}  analysed the chemo-dynamical and kinematic properties of the barred galaxies while \cite{2021A&A...650L..16F} studied the bar pattern speed. Table \ref{tab:AU_param} contains the most relevant morphological parameters used in this study.

Additionally, we characterize the strength of the bar by the maximum amplitude of the m=2 Fourier mode,  a common methodology used in the literature \citep{2002MNRAS.330...35A,2006ApJ...645..209D,2017MNRAS.469.1054A} . We directly obtain this parameter from the simulation as,
\begin{equation}
\label{ec:A_2}
A_{2} =max  \frac{\sqrt{a_{2}^{2}+b_{2}^{2}}}{a_{0}}
\end{equation}
where $a_{2}$ and $b_{2}$ are defined as,

\begin{equation}
\label{ec:a_2}
a_{2}(R) =  \sum_{i=0}^{N}m_{i}sin(2\theta_{i})
\end{equation}
 
\begin{equation}
\label{ec:b_2}
b_{2} (R)=  \sum_{i=0}^{N}m_{i}cos(2\theta_{i})
\end{equation}
being $m_{i}$  and $\theta_{i}$ the mass and azimuthal angle of a particle i at the the cylindrical radius R and N  the total number of particles at that radius. We use the maximum $A_{2}$ value within 10 kpc and set the limit value of 0.3 to define the formation of a strong bar. To prevent increases in $A_{2}$ caused by transient effects such as mergers we visually examine the maps to verify the formation of a bar. 
  
Auriga galaxies have evolved differently, some of them being in relative isolation with no major mergers since  z=2 while others have experienced multiple interactions and accretion events throughout their entire lives (a reduced number of galaxies are undergoing mergers at z=0). In this work we do not study AU-11, AU-29 and AU-30 because they exhibit strong signs of interactions in their z=0 surface brightness and perturbed SVE maps. We include AU-20, a galaxy that is currently undergoing a merger at z=0 because the infalling galaxy has not yet affected the central region. From the final sample of 27 galaxies almost three quarters (21 of them) are barred systems.

\subsection{SVE and central parts characterization}
\label{sec:SVE_characterizarion}

To analyze the SVE we study galaxies in a face-on configuration. To that end, we first place the angular momentum vector of stars younger than 3 Gyr within a sphere of radius 60 kpc  around the center of potential parallel to the vertical direction. These particles are mostly located in the young thin disc and thus, their angular momentum is less affected by external agents. This guarantees that the galaxy is properly rotated even when it is interacting with its satellites. Then, we project galaxies along the Z axis onto a square grid of 0.5 x 0.5 kpc pixels. The number of particles per pixel in the inner regions is quite large and no binning is required to reach a statistically significant number of particles compared to the analysis of the stellar disc \citep{2021arXiv210604187W}. We use a cylindrical restframe to characterize the SVE and we calculate the vertical, $\sigma_{z}$, radial, $\sigma_{r}$, and azimuthal, $\sigma_{\phi}$, velocity dispersions and the axial ratios $\sigma_{z}/\sigma_{r}$ and  $\sigma_{\phi}/\sigma_{r}$ in each pixel.


We aim to understand the effect of the bar on the SVE in the inner region of galaxies, and therefore, we limit our study to the region within, $R_{\rm{0,D}}$, the beginning of the disc-dominated region. Following \cite{2021arXiv210604187W} we characterize this parameter using the length of the bar, $L_{\rm{bar}}$, measured by \cite{2020MNRAS.491.1800B} and the disc scale length, $R_{\rm{d}}$, from \cite{AurigaProject}  in barred and unbarred galaxies, respectively. On the other hand, the effect of the bar on the stellar kinematics is less relevant in the central parts where bulges and nuclear discs have a larger impact. Therefore, we define our region of interest as the region between the effective radius of the bulge, $R_{\rm{B}}$ and $R_{\rm{0,D}}$. The effective radius of the bulge is taken from \cite{2020MNRAS.491.1800B} for barred galaxies and  \cite{AurigaProject} for unbarred systems.
 
\begin{table*}
	\centering
	\caption{Parameters of the Auriga galaxies at z=0. The columns are:(1) Galaxy name, (2) Hubble-Type, (3) Bulge effective radius, (4) Radius where the disc starts to dominate, (5) Bar strength, (6) Bar relative size, (7) Bar light fraction, (7) $\sigma_{z}$ categories from Sec. \ref{sec:profiles}.Galaxies with B/P bulges are indicated by $^{\ast}$}
	\label{tab:AU_param}
	\begin{tabular}{lccccccr} 
		\hline
		$Name$ & Hubble-Type& $R_{\rm{B}}$ (kpc) & $R_{\rm{0,disc}}$ (kpc) & $ f_{bar} $ & $ L_{bar}/R_{25} $ & Bar/T & Type $\sigma_{z}$ \\
		\hline
		AU-1 & SBb & 2.03 & 4.07 & 0.57 & 0.20 & 0.16 & B\\
		AU-2 & SBc & 1.22 & 8.99 & 0.59 & 0.24 & 0.07 & B\\
		AU-3 & Sb  & 1.49  & 7.26 & 0 & 0 & 0 &-\\
		AU-4 & Sbc  & 1.74  & 3.93 & 0 & 0 & 0 &-\\
		AU-5 & SBb  & 0.77 & 4.58 & 0.50 & 0.22 & 0.06 & B\\
		AU-6 & SBbc  & 0.98 & 5.43 & 0.42 & 0.21& 0.05 & A\\
		AU-7 & SBb  & 0.96 & 5.43 & 0.47 & 0.22 & 0.06 & A\\
		AU-8 & Sc  & 0.65 & 6.57 & 0 & 0& 0 & -\\
		AU-9 & SBb  & 0.98 & 6.45 & 0.76 & 0.34 & 0.15 & C\\
		AU-10 & SBa  & 0.85 & 6.45 & 0.75 & 0.40 & 0.33 & C\\
		AU-12 & SBab  & 0.85 & 3.73 & 0.60 & 0.20 & 0.06 & B\\
		AU-13$^{\ast}$ & SBa  & 0.97 & 5.43 & 0.59 & 0.35 & 0.29 & C\\
		AU-14 & SBb  & 0.81 & 5.09 & 0.65 & 0.20 & 0.09 & B\\
		AU-15 & Sbc  & 2.22 & 5.32 & 0 & 0 & 0.& -\\
		AU-16 & Sc  & 1.83 & 9.03 & 0 & 0 & 0& -\\
		AU-17$^{\ast}$ & SBa  & 1.26 & 5.43 & 0.59 & 0.34 & 19& C\\
		AU-18$^{\ast}$ & SBb  & 1.11 & 6.45 & 0.61 & 0.31 & 0.12& C\\
		AU-19 & Sbc  & 1.42 & 4.81 & 0 & 0 & 0 & -\\
		AU-20 & SBc  & 0.99 & 4.81 & 0.66 & 0.18 & 0.1& B\\
		AU-21 & SBb  & 1.02 & 3.39 & 0.41 & 0.14 & 0.04 & A\\
		AU-22$^{\ast}$ & SBa  & 0.80 & 6.28 & 0.56 & 0.47& 0.23 & C\\
		AU-23$^{\ast}$ & SBbc  & 1.31 & 9.5 & 0.69 & 0.38 & 0.11& C\\
		AU-24 & SBc  & 1.10 & 5.09 & 0.52 & 0.17 & 0.04 & B\\
		AU-25 & SBb  & 1.01 & 4.41 & 0.47 & 0.21 & 0.04 & A\\
		AU-26$^{\ast}$ & SBa  & 1.18 & 5.43 & 0.48 & 0.30 & 0.22 & C\\
		AU-27 & SBbc  & 0.87 & 5.77 & 0.69 & 0.22 & 0.08 & B\\
		AU-28 & SBa  & 0.90 & 7.46 & 0.82 & 0.43 & 0.36 & C\\
		\hline
	\end{tabular}
\end{table*}

\section{The SVE of the central region}
\label{sec:first_results}

In this section we use different approaches to explore the local properties of the SVE in the inner regions of Auriga galaxies. In \cite{2021arXiv210604187W} we focused on the SVE in the disc region of these galaxies and we showed the importance of analyzing its spatial distribution to fully characterize the underlying kinematic structure. In particular, we showed that similar global values can be obtained even if the SVE profiles and 2D maps present different behaviours. We expect that this behaviour is more pronounced in the inner regions due to the presence of non axisymmetric structures like the bar. Consequently, we study the SVE properties at both global and local levels.
 
\subsection{2D maps}
\label{sec:2Dmaps}

We first analyze the spatial 2D distribution of the SVE in the inner regions of Auriga galaxies. In Fig. \ref{fig:FIG1} we show the SVE 2D maps of seven galaxies selected from our sample. From left to right the maps show the V-band surface brightness and SVE velocity dispersions and axial ratios ($\sigma_{z}$, $\sigma_{r}$, $\sigma_{\phi}$, $\sigma_{z}/\sigma_{r}$ and  $\sigma_{\phi}/\sigma_{r}$, respectively). From top to bottom we find galaxies with different morphological features, e.g. an unbarred galaxy (AU-3), galaxies with weak bars (AU- 21 and AU-25), galaxies with strong bars (AU-24 and AU-27) and galaxies with a B/P (AU-17 and AU-23). The rest of the Auriga galaxies are shown in Appendix \ref{sec:appendix}. Inner and outer black circles indicate the boundaries of the region of interest.
We find that simulated galaxies present a complex SVE spatial distribution, particularly barred galaxies, and no general behaviour can be inferred even within each of the four categories. Nevertheless, we note that, on average, velocity dispersions increase inwards, except for some systems where the innermost regions have lower velocity dispersions (see  AU-4 in Fig. \ref{fig:FIG1app}).

Unbarred galaxies present the most homogeneous picture since the three velocity dispersions show inward increasing values in highly axisymmetric distributions.
Small scale differences are associated with stochasticity or dynamical agents that leave only a small imprint on the SVE such as the small spiral arms (e.g. AU-15 and AU-16 in Fig. \ref{fig:FIG1app}). Nevertheless, we notice that many galaxies present lower values in the innermost regions, specially in $\sigma_{\phi}$, e.g. AU-8 and AU-4. 
In the case of AU-8, the $\sigma_{\phi}$ map shows a modest non axisymmetric feature that suggest the presence of a very weak bar inside the bulge region. On the other hand, the low velocity dispersion region of AU-4 is formed after a recent merger that has injected gas into the galaxy. The SVE axial ratios maps are significantly different from one galaxy to the other and we do not find a general trend.

The panels of barred galaxies show a more complex SVE distribution where the velocity dispersion maps exhibit non-axisymmetric features that closely follow the bar structure. These features manifest as low velocity dispersion regions along the bar major axis within a hotter surrounding central region in the maps of the vertical and azimuthal SVE components. In galaxies with weak bars the low velocity dispersion region is confined to the central regions, extending only  few kilo-parsecs outside the bulge effective radius as in AU-25. In strongly barred systems like AU-24 this feature spans over the entire bar length. Similarly, the extent of the high velocity dispersion region around the bar is not  unique and  in some cases it may cover the entire region of study while in others it drops at intermediate radius. 
On the other hand, the maps of the radial velocity dispersion are similar to those of unbarred galaxies with larger $\sigma_{r}$ values inwards and no sign of axisymmetric features in most of the sample. Only two barred galaxies present an elongated high velocity dispersion feature along the bar (see the maps of AU-2 and AU-10 in Figs. \ref{fig:FIG1app} -\ref{fig:FIG2app}). The axial ratio maps reveal a similar trend where we find more oblate ellipsoids along the bar and larger axial ratios in the rest of the inner region. These features are mostly driven by the same low and high $\sigma_{z}$ and $\sigma_{\phi}$ features in the velocity dispersion maps.


\begin{figure*}
\centering
\includegraphics[width=0.99\textwidth]{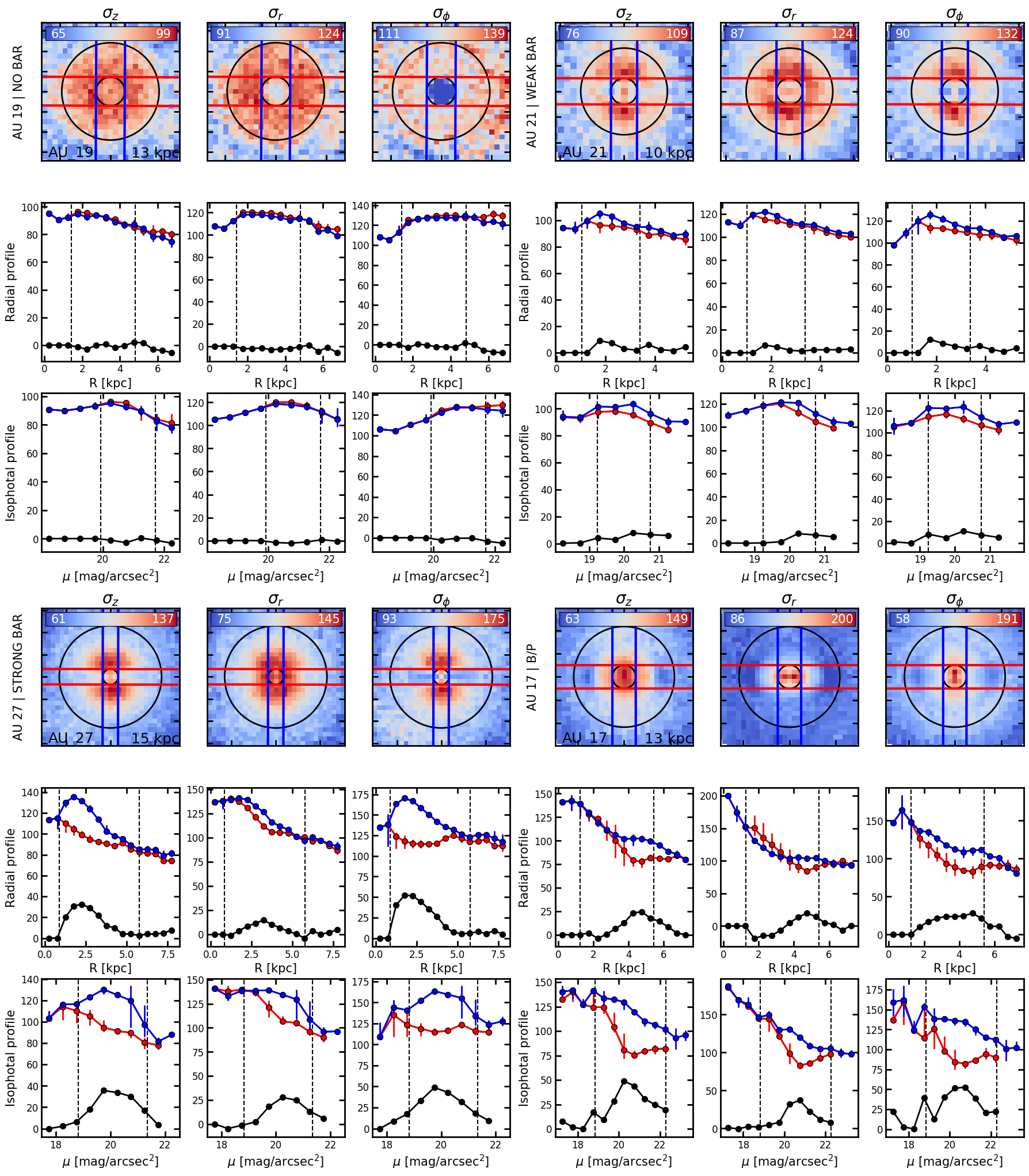}
\caption{Maps and profiles of an unbarred galaxy (top left),  a galaxy with a weak bar (top right), a galaxy with a strong bar (bottom left) and a barred galaxy with a B/P (bottom right). For each galaxy the panels show the projected 2D maps (top), radial profiles (middle) and isophotal profiles (bottom) of the vertical (left), radial (center) and azimuthal (right) velocity dispersions.  Inner and outer circles indicate the effective radius of the bulge and the beginning of the disc dominated region. Symbols and errorbars indicate the median, 16th and 84th value within each bin in the middle and bottom panels. Red and blue colours represent the results from the slits indicated in the maps along the bar major and minor axes. Black colour indicates the difference between these two regions. Vertical dashed lines mark the region of interest: position of the black circles for radial profiles and the average surface brightness at these positions for isophotal profiles. The area covered in this analysis is indicated (in kpc) at the bottom right of the left-most maps of each case.}
\label{fig:PROFILES}
\end{figure*}


Barred galaxies hosting B/P bulges present two main differences in their velocity dispersion maps with respect to the rest of barred galaxies. First, the low velocity dispersion region in the $\sigma_{z}$ maps does not cover the full extension of the bar. In fact, all six galaxies present a central high $\sigma_{z}$ region and lower values outwards. Further, the lowest values are found at the end of the bar e.g. AU-17. Secondly, all the B/P galaxies present the elongated high velocity dispersion feature along the bar in the $\sigma_{r}$ maps. Axial ratio maps are very similar to those of barred galaxies without BP despite the differences in the velocity dispersion maps.

In summary, the SVE spatial distribution in the inner regions of barred galaxies present  non-axisymmetric features in the velocity dispersion maps that closely match the bar location. These features present a qualitative dependence on the bar morphology since galaxies hosting stronger bars present more distinct regions with low $\sigma_{z}$ and $\sigma_{\phi}$ along the bar. Interestingly, all galaxies with B/P bulges present deviations from this general trend which emphasize the importance of this structure in the dynamics of galaxies. 

\subsection{Profiles}
\label{sec:profiles}

We now explore the 1D profiles to better analyze the features in the SVE maps in the previous section.  
To this end, we obtain the profiles along the bar major and minor axes using slits with a width twice the effective radius of the bulge which minimizes the bulge contribution in our analysis. Since the bar structure is easily detectable in the surface brightness map we focus on the isophotal profiles measured in bins of  0.5 mag/arcsec$^{2}$, following \cite{2015MNRAS.451..936S}. The results in the following sections are centered on the differences between these profiles in slits along the major and minor axes of the bar. Our choice of the isophotal profiles allows to better characterize the difference between barred and unbarred regions instead of averaging azimuthally  regions of the galaxy at the same distance from the center. Nevertheless, we have confirmed that similar results are obtained using radial profiles.

Fig. \ref{fig:PROFILES} shows the velocity dispersion maps and profiles of a galaxy without bar (AU-19 upper left), with a weak bar (AU-21, upper right), a strong bar (AU-27, bottom left) and a B/P (AU-17, bottom right). The panels show the projected 2D map (top), radial profiles (middle) and isophotal  profiles (bottom) of  $\sigma_{z}$ (left), $\sigma_{r}$ (center) and $\sigma_{\phi}$ (right) for each galaxy. Symbols and errorbars indicate the median and  the 16$^{th}$ and 84$^{th}$ percentiles in each bin along the bar major (red) and minor (blue) axes. The differences between minor and major axes (radial difference, $\Delta_{r}\sigma$, and isophotal difference, $\Delta_{\mu}\sigma$) are shown in black. Vertical dashed lines indicate the region of interest which correspond to the position of the black circles in the 2D maps for radial profiles and the average surface brightness at these positions for isophotal profiles.

For the unbarred galaxy we randomly selected the two perpendicular directions. Both radial and isophotal differences are negligible for the three SVE components. This behaviour is the same for the rest of unbarred galaxies. Only low level differences were obtained for some galaxies due to a combination of stochasticity and the presence of spiral arms close to the disc effective radius. This shows that there is no intrinsic difference in the inner spatial distribution of the SVE velocity dispersions. 

Barred galaxies exhibit differences as expected from the low velocity dispersion signatures in the 2D maps.  AU-21 is a weakly barred galaxy and the radial and isophotal profiles exhibit small differences in the three velocity dispersion components. The maps show that the characteristic low values along the bar (particularly in the $\sigma_{\phi}$ component) are confined in the bulge region and thus, we are not able to capture all the SVE differences associated to the bar. On the other hand, we can appreciate that the velocity dispersion is slightly larger in the perpendicular direction, around 10 km/s of difference. 

The other two barred galaxies (a strongly barred galaxy and a barred galaxy with B/P bulge) show a more complex scenario with significantly different velocity dispersion profiles. 
The plots of AU-27 reveal that all velocity dispersion profiles along the bar major axis present a smooth decreasing tendency with lower values than along the bar minor axis, leading to positive differences. The average behaviour is the same for radial (middle panels) and isophotal (bottom panels) profiles. Interestingly, we find that the vertical and azimuthal velocity dispersions have more similarities with each other, with a peak within the regions of interest along the minor axis, than with the radial component. In addition, we find that isophotal profiles better capture the influence of the bar in $\sigma_{r}$ compared to radial profiles where the differences are significantly lower.


The plots of AU-17 quantitatively show the behaviour previously described in Sec. \ref{sec:2Dmaps} where the velocity dispersion profile along the bar major axis presents a minimum close to the bar ends (both in the radial and isophotal profiles). In other words, the spatial behaviour of the velocity dispersions cannot be described with simply decreasing component, and in contrast to AU-27, the profiles along the bar minor axis present a smooth decreasing profile. We also find that isophotal differences are on average larger than radial differences and are always positive ($\sigma_{r}$ radial differences are negative in the inner regions due to high velocity dispersion features along the bar major axis). 



\begin{figure*}
\centering
\includegraphics[width=0.99\textwidth]{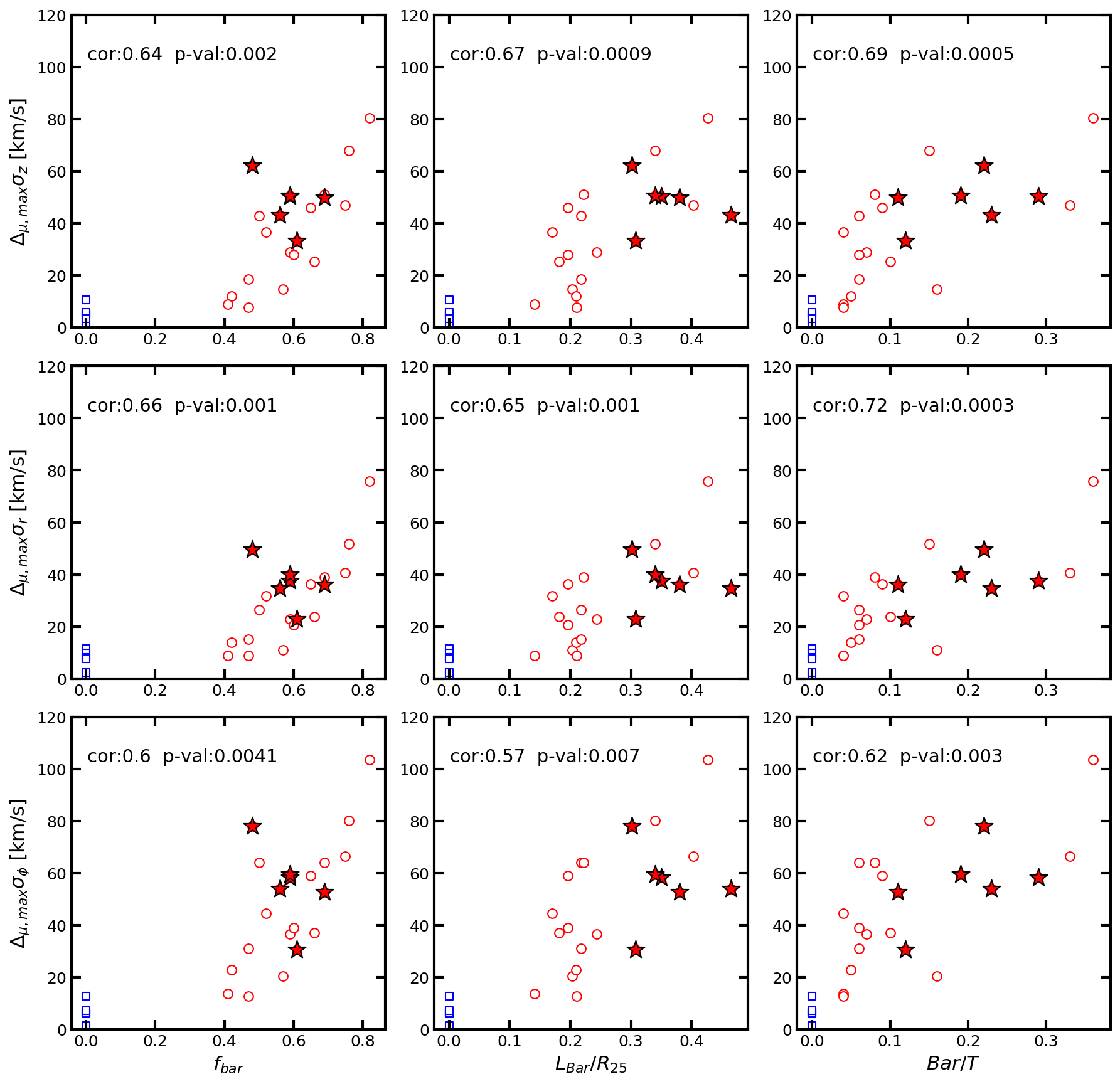}
\caption{The panels show the maximum isophotal difference of the vertical (top), radial (middle) and azimuthal (bottom) velocity dispersions as a function of the bar strength (left), relative size (center) and light fraction (right). Each panel shows the results for the full sample where each symbol represent a galaxy. Blue and red colours represent unbarred and barred galaxies, respectively. Filled red stars represent galaxies with a B/P. Numbers within each panel indicate the Spearman correlation coefficient and p-value of the barred sample.}
\label{fig:FIG8}
\end{figure*}

\begin{figure*}
\centering
\includegraphics[width=0.99\textwidth]{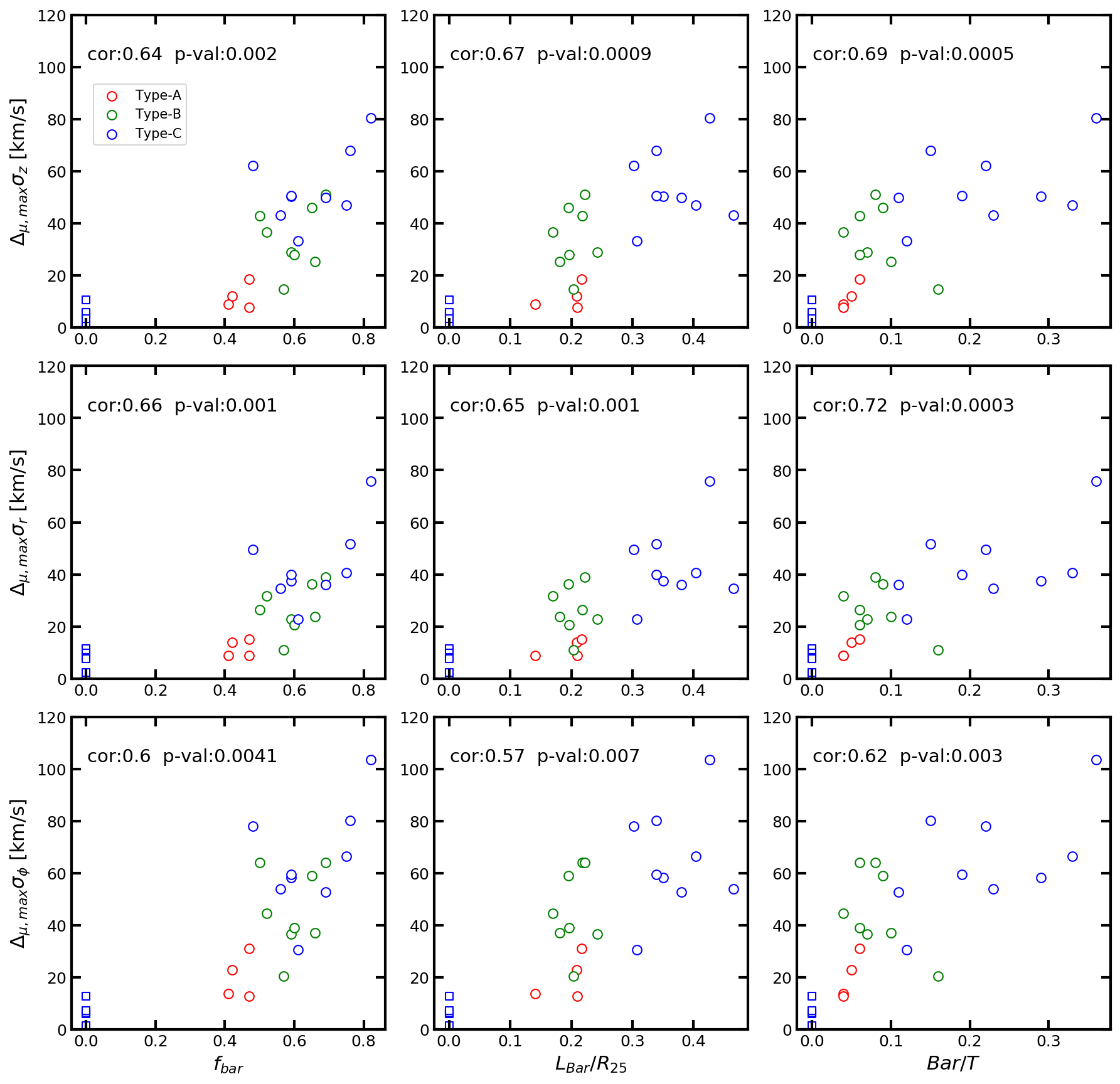}
\caption{Same as Fig. \ref{fig:FIG8}, color coding each galaxy according to the $\sigma_{z}$ classification from Sec. \ref{sec:profiles} }
\label{fig:FIG8_sigma_types}
\end{figure*}



The rest of barred galaxies in our sample present a combination of the features shown in these three examples which difficults the development of a complete classification of the SVE inner spatial distribution. Nevertheless, the results of such classification would be hard to test observationally except for the vertical velocity dispersion, which can be directly obtained form the line-of-sight kinematics in face on galaxies. This component is particularly important since it measures the vertical heating and is related to the thickening of disks. Thus, we divide galaxies into  three categories depending on the properties of their $\sigma_{z}$ profiles and differences.


\textbf{Type-A}:  the imprint of the bar is limited to the innermost region within the bulge and the differences in the region of interest are negligible, e.g. AU-21.

\textbf{Type-B}: the vertical velocity dispersion along the bar major axis smoothly decreases. Along the bar minor axis $\sigma_{z}$ increases until it peaks and then decreases to values similar to those along the bar major axis.

\textbf{Type-C}: the profiles along the bar present a multi component behaviour with a clear minimum close to the bar ends e.g. AU-17.

The results of this classification are shown in Table \ref{tab:AU_param}. We find that there are 4, 8 and 9 galaxies in each of these categories, and interestingly all B/P galaxies belong to the type-C one. The fact that not all type-C galaxies contain a B/P bulge suggests that these profiles are either obtained through two different mechanisms for galaxies with and without B/P bulge or that there is only one mechanisms that takes place in all the B/P galaxies in our sample.
 
The profiles of the SVE axial ratios of the barred sample do not exhibit the same differences as the individual dispersions and present a much homogeneous picture (not shown). This is in agreement with our previous results on the radial distribution of $\sigma_{z}/\sigma_{r}$ in the disc region where similar profiles were obtained with different velocity dispersion behaviours \citep{2021arXiv210604187W}.

To summarize, these results manifest that even though galaxies share some similitudes in the 2D maps, e.g. lower velocity dispersions along the bar major axis, a more in depth  analysis  is required to fully understand the behaviour of the SVE in the inner regions. In fact, these features can be very varied, from almost no imprint within the bar region to different combinations of increasing and decreasing profiles along the direction of the bar major and minor axes.

\section{Correlation with bar properties}
\label{sec:correlation}

In the previous section we found that stronger bars exhibit more clear differences between the velocity dispersion along the bar major and minor axes than smaller bars. Therefore, we now aim to quantify the connection between the bar morphology and the velocity dispersion features. To this end, we first choose a metric that allow us quantify the differences seen in Fig. \ref{fig:PROFILES}. In Sec. \ref{sec:profiles} we presented the radial and isophotal SVE profiles along the bar minor and major axes for a limited number of galaxies and we showed that while both radial and isophotal profiles have a similar behaviour, the latter provided larger differences. We also showed that in some cases, differences from radial profiles take negative values, particularly in the $\sigma_{r}$ component.
For these reasons, we will use the maximum isophotal difference, $\Delta_{\mu,max}\sigma$, i.e. the maximum difference between the isophotal profiles from the slits along the bar major and minor axes of the bar (see black lines in the bottom panels of Fig. \ref{fig:PROFILES}). This parameter allows us to better quantify the differences between the direction along and perpendicular to the bar. We have further ensured that similar results, although less significant, are obtained if we choose a different parameter (e.g. mean or median difference). Table \ref{tab:SVE_dif} contains the maximum differences from radial and isophotal profiles for the three velocity dispersions for all galaxies in our sample.

In terms of the average velocity dispersion in the region of interest, we find that both barred and unbarred galaxies have similar $\sigma_{r}$ and $\sigma_{\phi}$. In terms of the vertical component, $\sigma_{z}$, barred galaxies show on average larger values. This is somewhat expected since \cite{2016MNRAS.459..199G} showed that in Auriga galaxies bars can effectively heat stellar particles in the disc along the vertical direction, especially bars that experience buckling, in agreement with previous studies \citep{1984ApJ...282...61S,2006MNRAS.368..623M,2010ApJ...721.1878S,2015MNRAS.452.2367Y}. However, since not all galaxies have the same stellar mass and the number of unbarred galaxies in our sample is significantly lower we did not further explore the differences in the absolute velocity dispersion values between these two types of galaxies.

%

\begin{figure}
\centering
\includegraphics[width=0.5\textwidth]{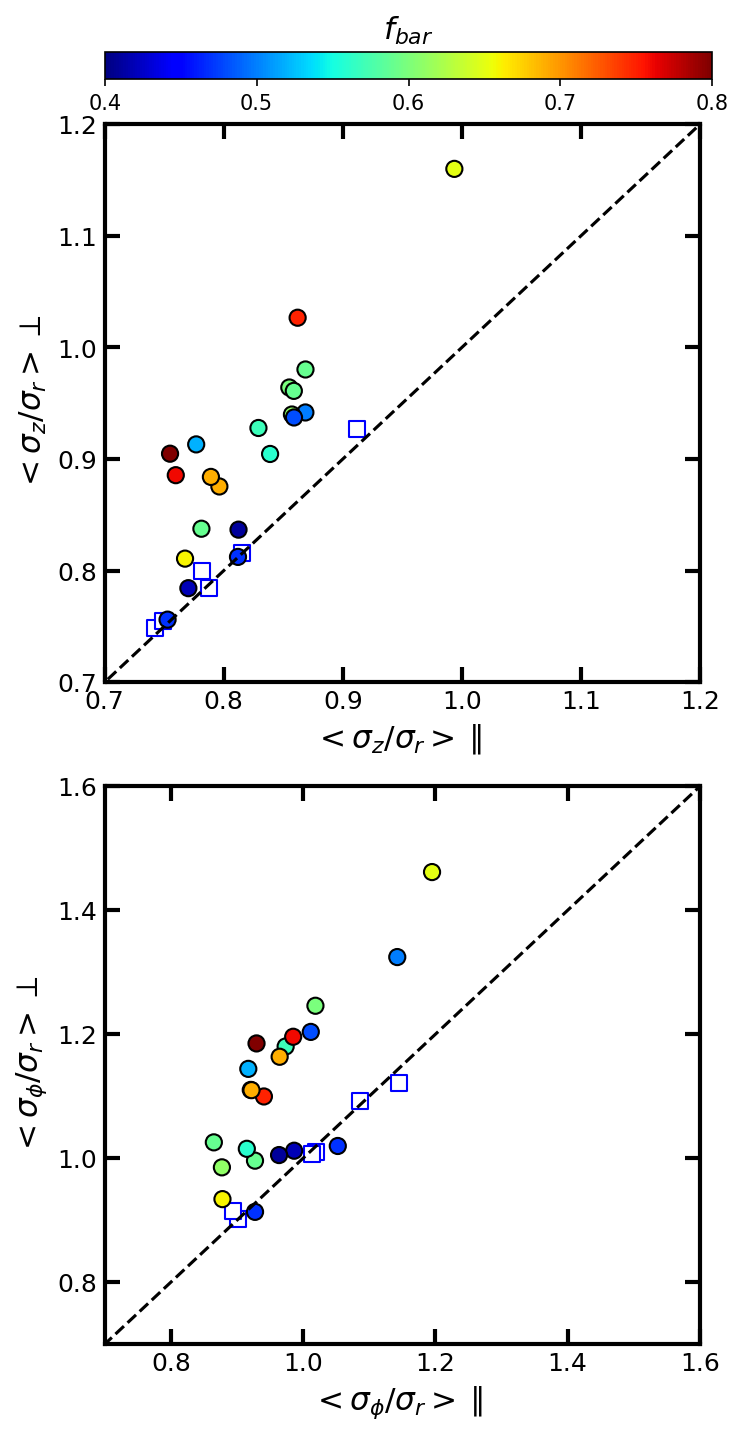}
\caption{Average inner $\sigma_{z}/\sigma_{r}$ (top) and $\sigma_{\phi}/\sigma_{r}$ (bottom) values along the bar minor axis (perpendicular direction ) as a function of the average values along the major axis (parallel direction). Blue squares represent unbarred galaxies and barred galaxies are indicated as circles that are color coded according to the bar strength. The black dashed lines represent the one to one correlation.}
\label{fig:FIG11}
\end{figure}

In Fig. \ref{fig:FIG8} we explore how the differences of the three velocity dispersions depend on different bar properties. From top to bottom the panels show the maximum isophotal differences of the vertical, radial and azimuthal velocity dispersions, as a function of the strength of the bar, $f_{bar}$ (left), the relative size of the bar, $L_{bar}/R_{25}$ (center) and the bar light fraction, Bar/L (right). Barred galaxies are represented with red circles and B/P with red stars. We use, unbarred galaxies (blue squares), as a control sample by setting the "bar properties"  equal to zero. Numbers within each panel indicate the Spearman correlation coefficient and p-value of the barred sample (red symbols).

\begin{figure*}
\centering
\includegraphics[width=0.9\textwidth]{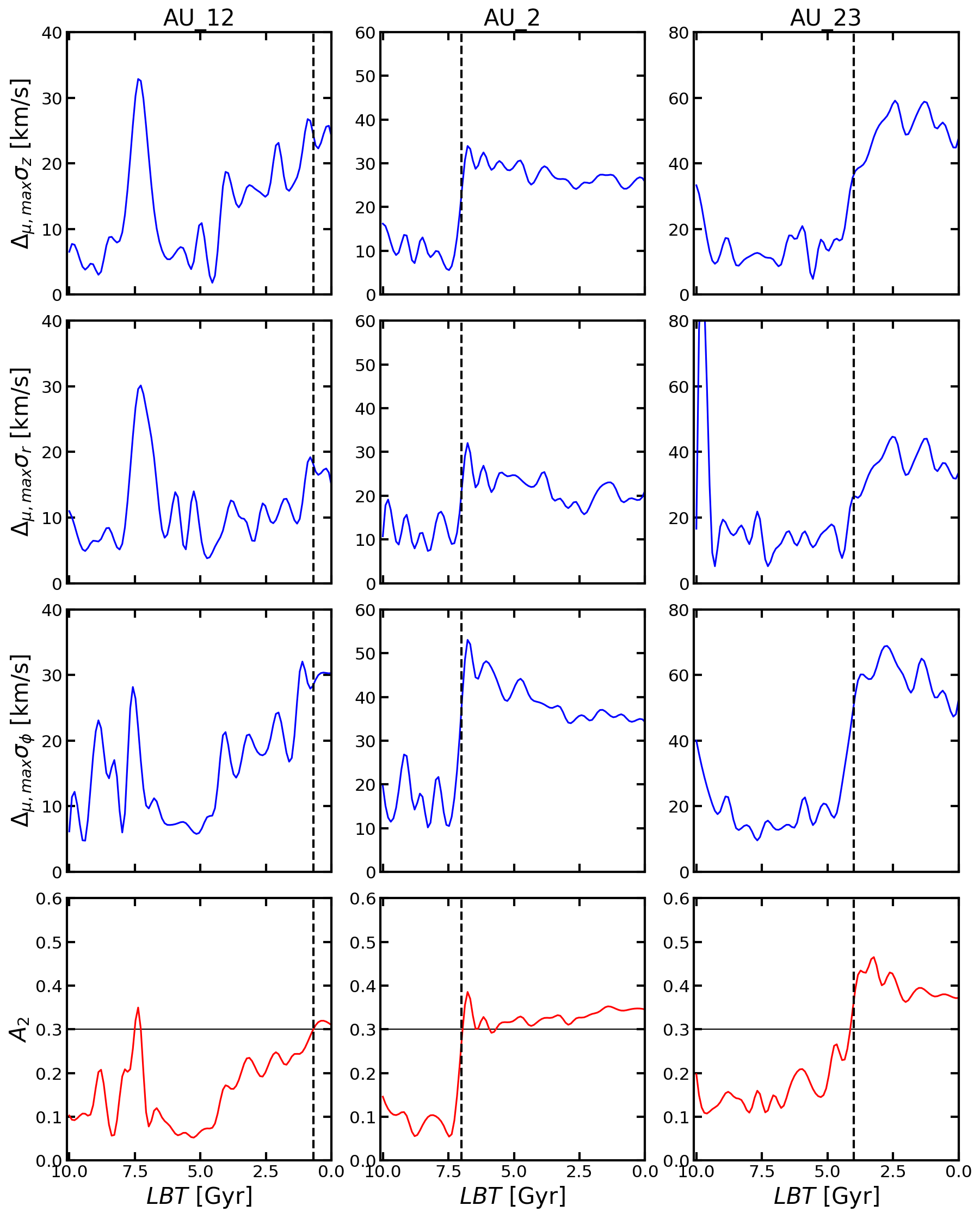}
\caption{From top to bottom the panels show  the time evolution of the vertical, radial and azimuthal velocity dispersion maximum isophotal differences, and the bar strength of  AU-12 (left), AU-2 (center) and AU-23 (right). The vertical dashed lines denote the formation of a strong bar with bar strength $A_{2}$ larger than the 0.3 threshold marked by the horizontal line in the bottom panels.}
\label{fig:FIG_EVOLTIME}
\end{figure*}

The panels show that there is a clear positive correlation between the velocity dispersion differences and the bar properties, i.e. galaxies with bars that present larger bar strength, relative size and light fraction exhibit larger differences in the three velocity dispersions. Unbarred galaxies present very small differences  as seen from their homogeneous 2D maps (see top maps in Fig. \ref{fig:FIG1}). 
The average difference of unbarred galaxies is about 10 km/s. This value could be considered a lower limit to assess whether the velocity dispersion differences of a barred galaxy are significant or not and therefore, they would be considered significant above this limit.  For example, in the left panels, we find that the SVE differences of barred galaxies without B/P bulge and with $f_{bar}$ around to 0.4 are close to this limit. In other words, the trend would not extend to lower values if we study galaxies with bar strength below 0.4. 

These differences present on average, significant and strong correlations with the bar parameters, showing an average Spearman coefficient around 0.7 and p-value below 0.01\% for the three velocity dispersions. We notice that the range of velocity dispersion differences is not the same for the three components being $\sigma_{\phi}$ the component that presents the largest values with a maximum of 103 km/s while $\sigma_{z}$ and $\sigma_{r}$  maximum values are around 80 and 75 km/s, respectively.

B/P galaxies exhibit intermediate differences that closely follow the general trend of the barred sample although there are two outliers in the left and middle panels. The first one is AU-26, which is the only system that presents larger differences than what would be appropriate for a galaxy with $f_{bar}$ of 0.5. This discrepancy is likely caused by a complex bar formation with significant changes in the $A_{2}$ parameter between 4 and 2 Gyrs ago, and the subsequent buckling instabilities which are present until present time. AU-26 is undergoing a secondary bucking phase (as seen by \citep{2020MNRAS.494.5936F}, and  \cite{2005AIPC..804..333A} and \cite{2006ApJ...637..214M} in other simulations). The second outlier is AU-22 a galaxy with a weak B/P that exhibits lower velocity dispersions than the general trend of barred galaxies in the $\Delta_{\mu,max}\sigma$-$L_{bar}/R_{25}$ panels. Remarkably, both galaxies follow the general trend if we use the Bar/T parameter to characterize the bar.

The 2D maps show that the most evident imprint of bars in the SVE spatial distribution are low velocity dispersion regions along the bar major axis. Therefore, it could be argued whether the trends seen in Fig. \ref{fig:FIG8} between the velocity dispersions and bar morphology proxies are genuine or rather, a connection between the velocity dispersion along the bar major axis ultimately drives these results. We repeated our analysis, calculating the average velocity dispersion in the slits along the bar major and minor axes, and studied their correlation with the strength of the bar. The results, (not shown), revealed only weak correlations with average \footnote{To compute these averages we consider the correlation coefficients and p-value of each direction, i.e. along the bar major and minor axes, and the three velocity dispersions,$\sigma_{z}$, $\sigma_{r}$ and $\sigma_{\phi}$} Spearman correlation coefficient of 0.2 and an average p-value of 40\%, thus, the null hypothesis cannot be discarded. These results pose a scenario where the morphology of the bar determines the extent of the velocity dispersion differences but does not affect the average value along the major and minor axes. We have further explored if these differences depend on other global properties of the galaxy, such as galaxy size or halo mass but the results (not shown) revealed that bar related parameters are the only ones that are correlated with the SVE differences.

Notwithstanding, we note that the parameters we use to characterize the morphology of the bar is an additional source of uncertainty. For example, in Fig.\ref{fig:FIG8} we use bar strengths measurements based on the ellipticity of the bar from \cite{2020MNRAS.491.1800B} which is only one of the multiple possibilities to characterize it. Alternatives would have been the use of radial to tangential forces, m=2 Fourier amplitudes or 2D kinematics. Fortunately, it has been shown that all these bar strength proxies are well correlated and we do not expect significant differences by choosing one over the others \citep{2015MNRAS.451..936S,2016A&A...587A.160D}. 

We have further explored if there is any connection between the $\sigma_{z}$ categories from Sec. \ref{sec:profiles} and the position in these panels.  Fig.\ref{fig:FIG8_sigma_types} shows the same results colour-coding each galaxy by the category they belong to (red, green and blue for type-A, B and C, respectively). The left panels show that type-A galaxies have the lowest values of bar strength and velocity dispersion isophotal difference. Galaxies in the other two categories show larger values in both quantities and are distributed in a very similar way. Conversely, the middle and right panels show a more interesting picture where the three categories occupy different regions. In particular, we find that there are two values that divides  type-C galaxies from the rest of the barred sample at $L_{bar}/R_{25}$ equal to 0.3 and Bar/T equal to 0.1. The type-A and type-B are reasonably divided in terms of their velocity dispersion differences. These results are particularly important because the classification was solely based on the isophotal profiles of the vertical velocity dispersion and did not take into account any morphological input.

The SVE axial ratio maps presented a much homogeneous picture than the three velocity dispersions maps with more oblate ellipsoids in the bar region than its surroundings. In Fig. \ref{fig:FIG11} we plot the average value of $\sigma_{z}/\sigma_{r}$ (top) and $\sigma_{\phi}/\sigma_{r}$ (bottom) along the bar minor axis ($\perp$) as a function of the average value along major axis ($\parallel$). Blue squares represent unbarred galaxies, barred galaxies are coloured according to the strength of the bar and the black dashed lines indicate the 1 to 1 correlation. As expected from the 2D maps, unbarred galaxies fall in the 1 to 1 correlation while the majority of barred galaxies lie above it. Following the previous results, most of the galaxies with low bar strength values present similar values to unbarred galaxies with almost the same SVE ratios along the two perpendicular directions. These galaxies belong to the type-A category. Surprisingly, there is no apparent correlation with the strength of the bar and the location of barred galaxies in these diagrams where they describe a well defined parallel sequence to the 1 to 1 correlation. Other bar morphology proxies and the type-B/C classification provide similar results. The average difference of the SVE axial ratios of barred galaxies between the minor and major axes is 0.08. 

In summary, our results show that bars do not affect the global properties of the SVE (velocity dispersions and axial ratios) but they determine the differences measured between the regions along the bar major and minor axes.

 \section{SVE time evolution}
\label{sec:time_evolution}

The results of the previous section showed a clear connection between the velocity dispersion differences and the bar properties at z=0. We now aim to understand what is the physical origin of these differences and how the formation and evolution of the bar affects the SVE. To that end, we study how the SVE evolve with time from z=2 to z=0, applying the same methodology described in Sec.\ref{sec:Methods} at different snapshots. In particular, we fixed the z=0 region of interest to study the SVE differences at higher redshifts. The average time step between snapshots is 0.16 Gyrs.


Fig. \ref{fig:FIG_EVOLTIME} shows from top to bottom the temporal evolution of the maximum isophotal difference of the vertical, $\sigma_{z}$, radial, $\sigma_{r}$ and azimuthal, $\sigma_{\phi}$, velocity dispersions, and the amplitude of the m=2 Fourier mode, $A_{2}$, as a function of the look back time. From left to right we find the results of AU-12, (type-B), AU-2, (type-B), and AU-23, (type-C with B/P). Vertical dashed lines indicate the time when we consider that the bar becomes strong i.e. $A_{2}$ > 0.3 and there is a clear bar in the surface brightness maps. This time does not always coincide with the bar formation since some bars have initial lower $A_{2}$ values and grow stronger over time (e.g. AU-12 where the bar formation began 5 Gyrs ago and it has steadily increased its strength) while others present large values since they are formed. Short-lived high $A_{2}$ values at high look back times are associated with mergers and satellite interactions.

The bottom left panel shows that the bar strength of AU-12 has steadily grown with time since the bar started to form 5 Gyrs ago and became strong 0.7 Gyrs ago. Interestingly, the upper panels show that the velocity dispersion isophotal differences follow a similar trend, increasing since the bar formation until present time. The vertical and azimuthal components present a time evolution that closely follows the evolution of $A_{2}$ while the the radial velocity dispersion shows a much weaker evolution.

Middle and right panels present a different formation scenario for AU-2 and AU-23 where both galaxies form a strong bar in a short period of time that causes an equally fast increase of the three velocity dispersion differences. Surprisingly, while there are fluctuations in small time scales we find that the differences of both galaxies present similar values in long periods of time, although a small decreasing trend can be appreciated mainly in the $\sigma_{\phi}$ component. We also find that the radial velocity dispersions present the lowest differences at all times. We notice that the bar of AU-23 experiences a buckling phase after the bar forms that causes the formation of a B/P bulge 2.5 Gyrs ago but no clear signatures of this process can be seen in the evolution of the velocity dispersion isophotal differences. 
In some situations low level differences can be associate with the formation of the BP bulge but the changes are similar to those experienced by galaxies without BP.

The rest of galaxies in our sample follow similar trends as these three examples. In galaxies where the bar experiences a more complex formation (e.g. substantial changes in bar strength during the formation or buckling of the bar) we find that the velocity dispersion differences tightly follow the $A_{2}$ evolution. In terms of the $\sigma_{z}$ classification, type-A galaxies present the weakest evolutions of the velocity dispersion isophotal differences and bar strength while type-B and type-C galaxies present similar behaviours. Lastly, the isophotal differences of the unbarred galaxies sample exhibit low-value flat curves with randomly distributed high values associated with mergers.

These plots confirm that the z=0 correlation between the bar properties and the velocity dispersion differences is caused by the bar since its formation induces the differences in the velocity dispersion. Further, they reveal that these differences remain fairly constant since the formation of the bar. Surprisingly, our results indicate that there is no apparent connection between the velocity dispersion differences and the B/P formation.

 \section{Discussion}
\label{sec:discussion}

Previous sections show that bars affect the kinematics of stars in the inner regions of galaxies inducing non axisymmetric features in the z=0 velocity dispersion maps that coincide with the bar light distribution. Further, we find that velocity dispersion differences between the minor and major bar axes are larger for galaxies with stronger bars. We now explore whether there is an observational counterpart to our findings and if other simulations provide similar results.

Thanks to the development of Integral Field Units we are able to fully study the 2D kinematics of galaxies. From the first works from the SAURON survey \citep{2007MNRAS.379..418C} to the latest results from ManGA \citep{2018MNRAS.477.4711G} and SAMI \citep{2021arXiv210906189V,2021MNRAS.505.3078V} our understanding of the kinematic properties of galaxies is steadily expanding. However, many of these works have focused on global properties galaxies and the number of dedicated works to study in detail the differences caused by the bar in late-type systems is scarce.
\cite{2015MNRAS.451..936S} is one of the first works to study both the radial and isophotal velocity dispersion profiles and 2D maps of 16 galaxies from the BALROG sample. These galaxies are not face-on and thus it is difficult to compare the line-of-sight velocity dispersion from BALROG with intrinsic velocity dispersions from AURIGA. Nevertheless, the profiles presented in their Appendix C show that many galaxies are compatible with type-A and type-C classifications although the 2D features are not so clearly seen in the maps. 

The velocity dispersion signatures from type-B galaxies have been observed in the velocity dispersion maps of inner bars in double barred galaxies.  \cite{2008ApJ...684L..83D,2012MNRAS.420.1092D,2019MNRAS.484..665D} have shown that these systems present low velocity dispersions along the inner-bar major axis and larger values with a peak in the perpendicular direction, features that have been coined as $\sigma$-hollows \citep{2008ApJ...684L..83D} and $\sigma$ humps \citep{2016ApJ...828...14D}, respectively. \cite{2008ApJ...684L..83D} proposed that these features were caused by a contrast between the cold orbits of stars from the inner bar and the high velocity dispersion from the bulge component. N-body simulations have shown that these features can also be detected in the presence of B/P bulge formed by the buckling of the outer bar \citep{2016ApJ...828...14D}. On the other hand, by applying Jeans equations to a bar + disc system \cite{2017ApJ...836..181D} have shown that there is no need for a hot component to obtain a $\sigma$ hollow+hump pattern in the vertical velocity dispersion since these features appear naturally as a dynamical response of stars to the presence of a massive vertically thin bar.

Bars in Auriga galaxies extend far beyond the bulge effective radius and the contribution of the bulge hot orbits to our velocity dispersion features should be minimal. Nevertheless, we repeated our analysis applying a vertical cut to stellar particles beyond two bulge effective radius from the plane of the disc which also discards halo stars. The results (not shown) provide velocity dispersion maps with the same features and similar correlations between the minor to major velocity dispersion differences and bar morphology proxies, thus discarding this possibility.

From the point of view of simulations, the number of works exploring the kinematics of galaxies in a cosmological environment have increased in the last years. Nevertheless, they have focused mostly on global kinematic properties and do not explore in detail the questions proposed here \citep[e.g.][]{2017MNRAS.464.3850L,2018MNRAS.476.4327L,2018MNRAS.473.4077P,2018MNRAS.480.4636S,2020arXiv201208060L,2020MNRAS.494.5652W,2022arXiv220108406F}.
Interestingly, \cite{2018MNRAS.473.4077P} presented the analysis of the kinematics of galaxies from IllustrisTNG50 (a high resolution cosmological volume simulation) and showed that the velocity dispersion maps of face-on disc galaxies (see Fig. 4 in their work) present similar low velocity dispersion features along the bar major axis to those presented in this work. This opens the possibility to study the role of the bar in the SVE  in a different environments with an independent and larger dataset. 

Our results, motivate more theoretical studies aimed to understand how the kinematics and morphology of bars coevolve to form the $\sigma$-hollows at z=0 and how they are connected to the evolution of the rest of the components. In addition, our search of the literature indicate that more integral-field observations of low inclination single barred galaxies are required to understand how common these features are in our Universe. Furthermore, these works would provide observational constrains that will allow to determine wether the intrinsic kinematic properties of simulated galaxies are well reproduced by currently used models.

We have shown that the velocity dispersion differences of the three components are well stablished since the formation of the bar but is not clear how do different velocity dispersion distributions are formed. To answer these questions we could explore if there is any signature at z=0 that could help distinguish the formation mechanism and evolution of the bar by studying the SVE of different stellar populations. Preliminary results show that on average young(old) stellar populations have colder(hotter) kinematics in agreement with a secular evolution scenario. However, we did not find any general behaviour in the velocity dispersion maps. Young and old populations present complex spatial variations that deviate from the patterns obtained for all the particles to a greater or lesser extent for each galaxy depending on the individual dynamical evolution and the star formation history in the bar region. To fully characterize the SVE for different populations in the inner regions exceeds the scope of this paper and we will focus on this topic in future works.

 \section{Conclusions}
\label{sec:conclusions}

We have analyzed a set of 27 high resolution cosmological zoom-in simulations of Milky-Way mass late-type galaxies from the Auriga sample to study the spatial distribution of the Stellar Velocity Ellipsoid (SVE). In particular, we studied the impact of the bar on the SVE properties by analyzing the 2D maps of the three velocity dispersions and their ratios (e.g. $\sigma_{z}$ , $\sigma_{r}$, $\sigma_{\phi}$, $\sigma_{z}$/$\sigma_{r}$ and $\sigma_{\phi}$/$\sigma_{r}$.)

The 2D velocity dispersion maps show a variety of features in the different components that depend on the presence of structures such as bars or B/P bulges. The most common feature is a low velocity dispersion region along the bar in the $\sigma_{z}$ and $\sigma_{\phi}$ maps which is more evident in galaxies with strong bars. Conversely, SVE axial ratio maps are very similar with no strong dependence on the bar properties.

A more in depth analysis of the isophotal profiles exhibit different behaviours along the bar major and minor axes. We classified galaxies into three categories attending to the vertical velocity dispersion profiles which could be confirmed observationally. In particular, galaxies can exhibit similar profiles along the two direction with almost no differences (type-A), smooth decreasing profile along the bar and increasing profile with a peak in the perpendicular direction (type-B) and a multi component decreasing profile along the bar major axis with a minimum close to the bar ends (type-C).

The minor-major velocity dispersion differences obtained from the isophotal profiles are well correlated with structural properties for the three SVE components. In other words, the velocity dispersions differences are more significant in barred galaxies with larger bar strength, relative size and light fraction. On the contrary, the correlation between the velocity dispersions and the bar properties is significantly weaker.In terms of the SVE axial ratios, the bar regions present more oblate ellipsoids but there is no apparent connection with the bar  properties.

The time evolution of the SVE and the bar strength confirm that the velocity dispersion differences seen at z=0 are caused by the bar formation. In bars that slowly grow over time the velocity dispersion differences follow an equally steady evolution, while strong bars that are formed in short time intervals present a rapid increase in the differences that remain very stable over large periods of time. 

Future works using cosmological volume simulations such as IlustrisTNG50 \citep{2018MNRAS.473.4077P} will help us understand how do these findings adapt to galaxies in different environments and with different masses.

\section*{Acknowledgements}
D. W-M, F. P and J. F-B acknowledge support through the RAVET project by the grant PID2019-107427GB-C32 from the Spanish Ministry of Science, Innovation and Universities (MCIU), and through the IAC project TRACES which is partially supported through the state budget and the regional budget of the Consejer\'ia de Econom\'ia, Industria, Comercio y Conocimiento of the Canary Islands Autonomous Community. RG acknowledges financial support from the Spanish Ministry of Science and Innovation (MICINN) through the Spanish State Research Agency, under the Severo Ochoa Program 2020-2023 (CEX2019-000920-S)



\section*{Data Availability}
The data underlying this article will be shared on reasonable request to the corresponding author.

\bibliographystyle{mnras}
\bibliography{SVE_BAR} 

\appendix
\section{2D SVE maps}
\label{sec:appendix}

\begin{figure*}
\centering
\includegraphics[width=0.99\textwidth]{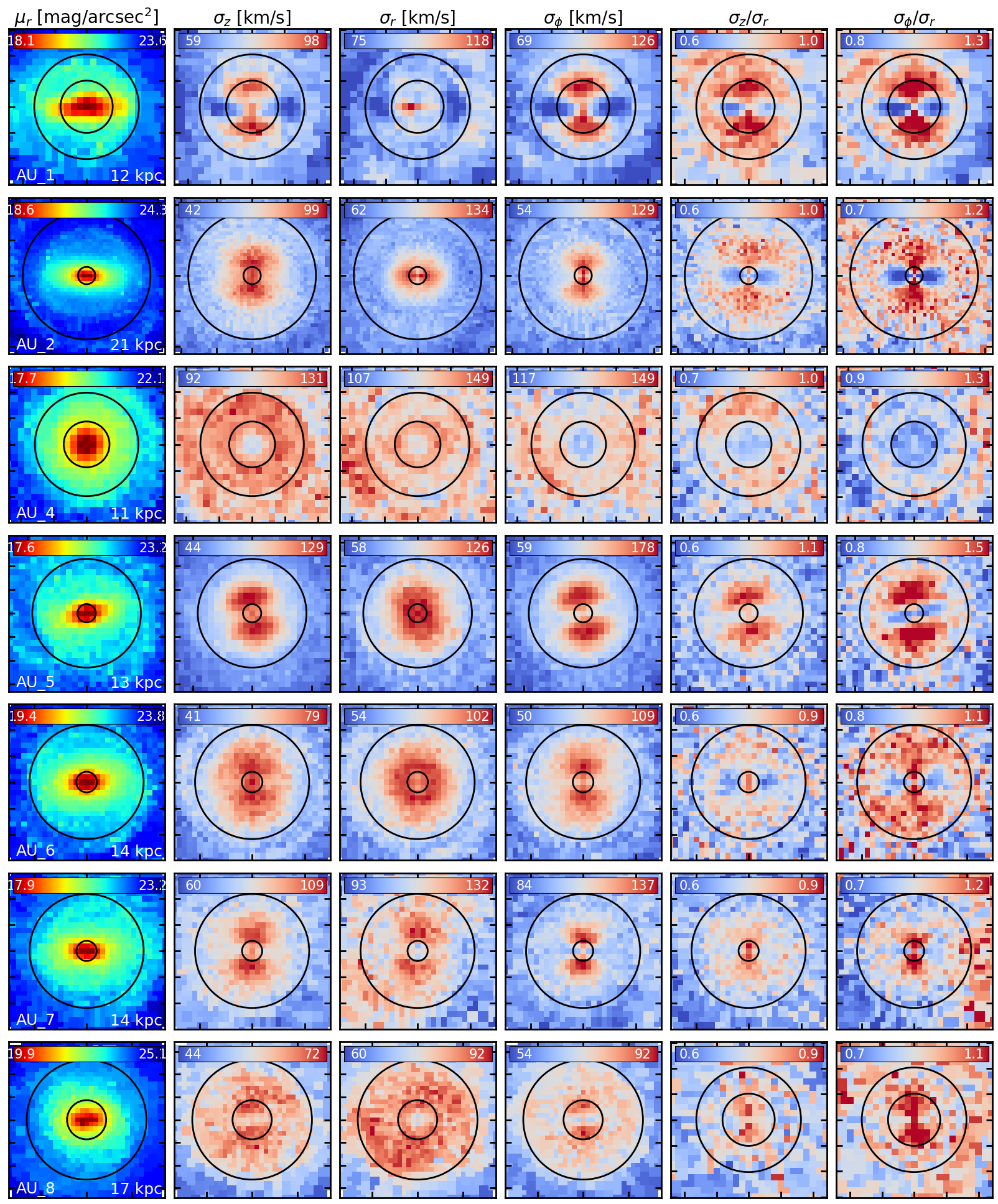}
\caption{Continuation of Fig.\ref{fig:FIG1} for the rest of Auriga galaxies.}
\label{fig:FIG1app}
\end{figure*}

\begin{figure*}
\centering
\includegraphics[width=0.99\textwidth]{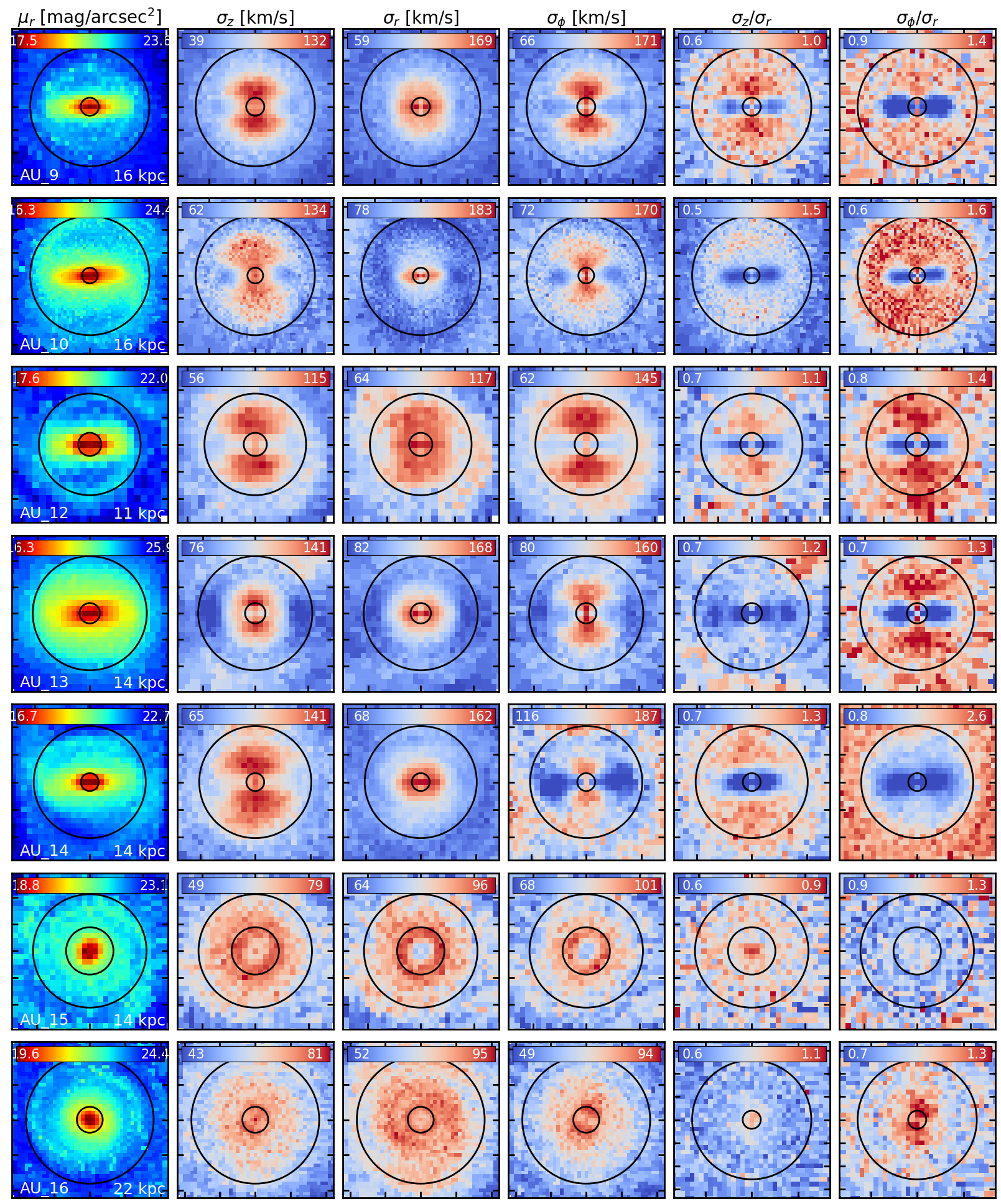}
\caption{Continuation of Fig.\ref{fig:FIG1} for the rest of Auriga galaxies.}
\label{fig:FIG2app}
\end{figure*}

\begin{figure*}
\centering
\includegraphics[width=0.99\textwidth]{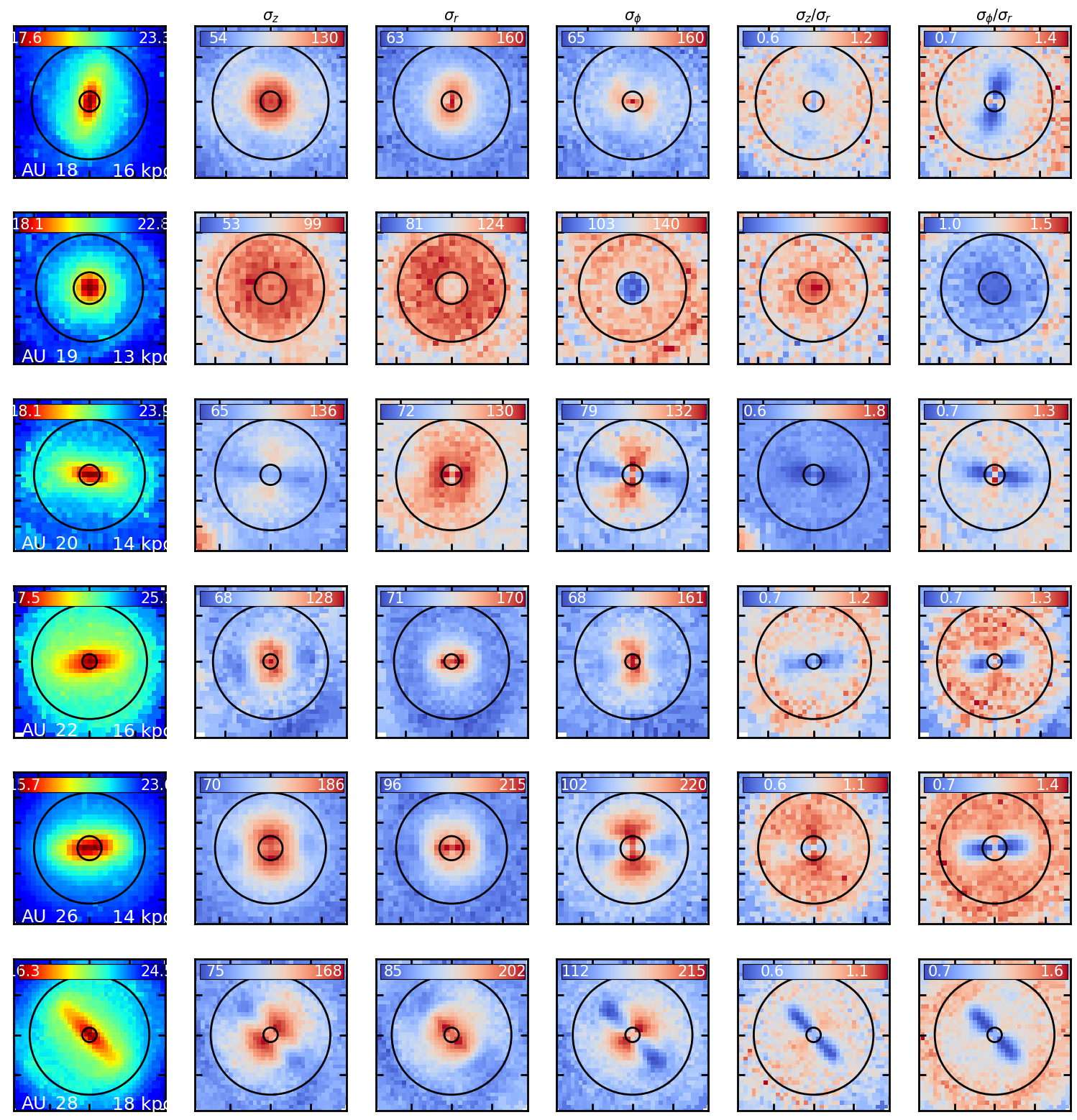}
\caption{Continuation of Fig.\ref{fig:FIG1} for the rest of Auriga galaxies.}
\label{fig:FIG3app}
\end{figure*}

\section{SVE isophotal and radial differences values}
\label{sec:appendix2}

\begin{table*}
	\centering
	\caption{Maximum radial and isophotal differences of Auriga galaxies at z=0. The columns are:(1) Galaxy name, (2)-(4) Maximum radial difference of the vertical, radial and azimuthal velocity dispersion, (5)-(7) Maximum isophotal difference of the vertical, radial and azimuthal velocity dispersion}
	\label{tab:SVE_dif}
	\begin{tabular}{lccccccr} 
		\hline
		$Name$ & $\Delta_{r,max}\sigma_{z}$ [km/s]& $\Delta_{r,max}\sigma_{r}$ [km/s]& $\Delta_{r,max}\sigma_{\phi}$ [km/s] & $\Delta_{\mu,max}\sigma_{z}$ [km/s]& $\Delta_{\mu,max}\sigma_{r}$ [km/s]& $\Delta_{\mu,max}\sigma_{\phi}$ [km/s] \\
		\hline
		AU1 & 19.2 & 14.7 & 19.9 & 14.8 & 11.0 & 20.6\\
		AU-2 & 13.7 & 13.8 & 26.2 & 28.9& 22.7 & 36.7\\
		AU-3 & 6.4 & 11.0 & 12.2 & 10.6 &11.6 & 12.8\\
		AU-4 & 8.6 & 12.8 & 11.6 & 5.4 & 9.8 & 6.5\\
		AU-5 & 26.8 & 12.7 & 58.9 & 42.8 & 26.5 & 64.1\\
		AU-6  & 8.9 & 5.7 & 19.5 & 12.0 & 14.0 & 22.9\\
		AU-7 & 17.2 & 10.8 & 27.3 & 18.6 & 15.1 & 31.0\\
		AU-8 & 5.1 & 13.5 & 9.3 & 4.9 & 8.2 & 6.2\\
		AU-9 & 38.1 & 14.1 & 63.2 & 68.1 & 51.6 & 80.1\\
		AU-10 & 36.9 & 24.9 & 45.8 & 46.9 & 40.6 & 66.6\\
		AU-12 & 25.9 & 12.0 & 44.0 & 28.0 & 20.7 & 39.1\\
		AU-13$^{\ast}$ & 24.5 & 25.1 & 42.4 & 50.4 & 37.6 & 58.4\\
		AU-14 & 28.0 & 14.2 & 45.7 & 46.0 & 36..3 & 59.0\\
		AU-15 & 3.8 & 4.3 & 4.0 & 1.2 & 1.5 & 1.4\\
		AU-16 & 7.0 & 11.2 & 5.7 & 5.9 & 7.5 & 6.5\\
		AU-17$^{\ast}$  & 19.3 & 19.4 & 18.4 & 50.6 & 39.9 & 59.5\\
		AU-18$^{\ast}$ & 8.3 & 14.7 & 18.2 & 33.3 & 22.9 & 30.7\\
		AU-19 & 6.3 & 8.9 & 11.0 & 3.3 & 2.3& 7.4\\
		AU-20 & 19.8 & 11.7 & 30.2 & 25.4 & 23.7 & 37.2\\
		AU-21 & 9.8 & 6.5 & 14.2 & 9.0 & 8.7 & 13.8\\
		AU-22$^{\ast}$  & 26.7 & 29.3 & 30.1 & 43.2 & 34.7 & 54.0\\
		AU-23$^{\ast}$ & 16.3 & 18.9 & 35.1 & 50.0 & 36.0 & 52.8\\
		AU-24 & 33.6 & 12.8 & 50.5 & 36.7 & 31.6 & 44.6\\
		AU-25 & 11.1 & 14.4 & 15.9 & 7.7 & 8.8 & 12.9\\
		AU-26$^{\ast}$ & 24.4 & 17.1 & 64.8 & 62.3 & 49.5 & 78.0\\
		AU-27 & 31.2 & 9.4 & 53.5 & 51.1 & 39.0 & 64.2\\
		AU-28 & 31.6 & 38.3 & 59.0 & 80.4 & 75.7 & 103.7\\
		\hline
	\end{tabular}
\end{table*}




\bsp	
\label{lastpage}
\end{document}